\documentclass[twoside,british,a4paper,twocolumn]{article}

\usepackage{amsmath}
\usepackage{amsfonts}
\usepackage{amssymb,upref}
\usepackage{tgheros} 
\usepackage{tgtermes} 
\usepackage{anyfontsize}
\usepackage{xcolor}
\usepackage{graphicx}
\usepackage{enumitem}
\usepackage{subfig}
\usepackage{siunitx}
\usepackage{empheq}

\usepackage[utf8]{inputenc}
\usepackage[T1]{fontenc}

\usepackage{textcomp}

\usepackage{isodate}
\cleanlookdateon

\usepackage{lettrine}
\pdfmapfile{=montserrat.map} 

\usepackage{lipsum}

\usepackage{newtxmath}

\usepackage{geometry}
\usepackage{fancyhdr}

\fancypagestyle{firstpagestyle}
\usepackage{titlesec}
\titleformat{\section}[block]{\bfseries\upshape\sffamily\boldmath}{}{0.em}{}
\titlespacing*{\section}{0pt}{0.8em plus 0ex minus 0ex}{0em plus 0.ex}

\titleformat{\subsection}[block]{\bfseries\upshape\sffamily\footnotesize}{}{0.em}{}
\titlespacing*{\subsection}{0pt}{0.5em plus 0ex minus 0ex}{0em plus 0.ex}

\usepackage{titling}
\pretitle{\noindent\begin{minipage}{\textwidth} \raggedright \Huge \bfseries\upshape\sffamily\boldmath
\color[rgb]{0,0,0.8}
}
\posttitle{\end{minipage} \vskip 0.6em}
\preauthor{\noindent \begin{minipage}{\textwidth} \large\mdseries\sffamily}
\postauthor{
  \end{minipage} 
  \vskip 0.4em
  {
   \sffamily
   \color{black!80}
   \noindent \small
   \address
   \par 
   \authoremail
   }
   \par \vskip 0.4em
  }
\predate{ \noindent \hspace{-0.5em} \sffamily\footnotesize}
\postdate{\vskip 0.0em}

\usepackage{caption}
\DeclareCaptionFont{cfs}{\fontsize{8.5}{10.25}\selectfont}
\DeclareCaptionLabelSeparator{vline}{\;|\;}
\newcommand{\figurecaption}[2]{\caption[#1]{\textbf{#1.} #2}}

\newcommand{\orcid}[1]{%
  \href{%
    https://orcid.org/#1%
  }{%
   \,\protect\includegraphics[width=8pt]{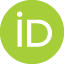}%
  }%
}

\graphicspath{{Figures/}}

\usepackage{tcolorbox}
\definecolor{abstractboxcolor}{cmyk}{0.1,0,0,0}
\newtcolorbox{abstractbox}{
  arc=0pt,
  boxrule=0pt,
  colback=abstractboxcolor,
  boxsep=0.5em,
  left=0pt, right=0pt, bottom=0pt, top=0pt,
  width=\columnwidth
}
\makeatletter
 \def\@textbottom{\vskip \z@ \@plus 1pt}
 \let\@texttop\relax
\makeatother
\renewenvironment{abstract}{
   \noindent
   \begin{minipage}{\textwidth}
   \upshape\sffamily \bfseries
   \fontsize{9}{11.5}\selectfont
  }{
   \end{minipage} 
   \vskip 2.0em
  }

\usepackage[numbers,sort&compress]{natbib}

\makeatletter \def\NAT@def@citea{\def\@citea{\NAT@separator\,}} \makeatother 
\usepackage{etoolbox}
\apptocmd{\sloppy}{\hbadness 10000\relax}{}{}

\newcommand{\citer}[1]{Ref.~\citealp{#1}}

\newcommand{\reffig}[1]{Fig.~\ref{#1}}

\usepackage[%
  bookmarks=true,
  colorlinks,
  linkcolor=blue,
  urlcolor=blue,
  citecolor=blue,
  plainpages=false,
  pdfpagelabels,
  final,
  breaklinks=true,
]{hyperref}
\hypersetup{
  pdftitle={Knotting fractional-order knots with the polarization state of light}, 
  pdfauthor={Emilio Pisanty, Gerard J. Machado, Verónica Vicuña-Hernández, Antonio Picón, Alessio Celi, Juan P. Torres and Maciej Lewenstein}
}

\title{Knotting fractional-order knots with the polarization state of light}
\newcommand\shorttitle{Knotting fractional-order knots with the polarization state of light}

\author{%
Emilio Pisanty\orcid{0000-0003-0598-8524}\textsuperscript{1$*$}, %
Gerard J. Machado\orcid{0000-0002-0841-2195}\textsuperscript{1}, %
Verónica Vicuña-Hernández\orcid{0000-0002-6926-8772}\textsuperscript{1}, %
Antonio Picón\orcid{0000-0002-6142-3440}\textsuperscript{1$\!$,$\:\!$2}, \\%
Alessio Celi\orcid{0000-0003-4939-084X}\textsuperscript{1$\!$,$\:\!$3,$\:\!$4,$\:\!$5}, %
Juan P. Torres\orcid{0000-0002-4454-6676}\textsuperscript{1$\!$,$\:\!$6} and %
Maciej Lewenstein\orcid{0000-0002-0210-7800}\textsuperscript{1$\!$,$\:\!$7}%
}
\newcommand\shortauthor{E. Pisanty, G.J. Machado, V. Vicuña-Hernández, A. Picón, A. Celi, J.P. Torres and M. Lewenstein}

\newcommand\address{%
\fontsize{8}{11}\selectfont
\setlength{\baselineskip}{1.2em}
\begin{enumerate}[itemsep=-0.5ex, labelsep=0.1em, leftmargin=0.7em, label=\textsuperscript{\arabic*} ]
\item ICFO -- Institut de Ciencies Fotoniques, The Barcelona Institute of Science and Technology, Av. Carl Friedrich Gauss 3, 08860 Castelldefels (Barcelona), Spain
\item Departamento de Química, Universidad Autónoma de Madrid, 28049, Madrid, Spain
\item Center for Quantum Physics, University of Innsbruck, Innsbruck, Austria
\item Institute for Quantum Optics and Quantum Information, Austrian Academy of Sciences, Innsbruck, Austria 
\item Departament de Física, Universitat Autònoma de Barcelona, E-08193 Bellaterra, Spain
\item Department of Signal Theory and Communications, Universitat Politecnica de Catalunya, Barcelona, Spain
\item ICREA, Passeig de Lluís Companys, 23, 08010 Barcelona, Spain
\end{enumerate}
}
\newcommand\authoremail{\noindent $^*$emilio.pisanty@icfo.eu}

\date{
10 June 2019
\\[1mm]
This is the Submitted Manuscript for \href{https://doi.org/10.1038/s41566-019-0450-2}{\color[rgb]{0,0,0.4}\textit{Nature Photonics} doi:10.1038/s41566-019-0450-2 (2019)}, \href{https://arxiv.org/abs/1808.05193}{\color[rgb]{0,0,0.4}arXiv:1808.05193}.
}

\usepackage{physics}

\renewcommand{\Re}{\operatorname{Re}}

  \newcommand{\vbe}{\vb{E}}
  \newcommand{\vbr}{\vb{r}}
  
  \newcommand{\tve}{\tilde{\vb{E}}}

\newcommand{\ue}[1]{\hat{\vb{e}}_{#1}}

\newcommand{\uu}{\hat{\vb{u}}}

\newcommand{\noheadersection}[1]{%
  \par\refstepcounter{section}
  \sectionmark{#1}
  \addcontentsline{toc}{section}{\protect\numberline{\thesection}#1}
}

\begin{document}

\twocolumn[
\begin{@twocolumnfalse}

\maketitle
\thispagestyle{firstpagestyle}

\begin{abstract}
The fundamental polarization singularities of monochromatic light are normally associated with invariance under coordinated rotations: symmetry operations that rotate the spatial dependence of an electromagnetic field by an angle~$\boldsymbol \theta$ and its polarization by a multiple $\boldsymbol {\gamma \theta}$ of that angle. These symmetries are generated by mixed angular momenta of the form $\boldsymbol{J_\gamma = L+\gamma S}$ and they generally induce Möbius-strip topologies, with the coordination parameter $\boldsymbol \gamma$ restricted to integer and half-integer values. 
In this work we construct beams of light that are invariant under coordinated rotations for arbitrary $\boldsymbol \gamma$, by exploiting the higher internal symmetry of `bicircular' superpositions of counter-rotating circularly polarized beams at different frequencies. We show that these beams have the topology of a torus knot, which reflects the subgroup generated by the torus-knot angular momentum $\boldsymbol{J_\gamma}$, and we characterize the resulting optical polarization singularity using third-and higher-order field moment tensors, which we experimentally observe using nonlinear polarization tomography.
\end{abstract}

\vspace{-2mm}

\end{@twocolumnfalse}
]

\noheadersection{Introduction}
\lettrine[lines=3, lhang=0.15]{T}{\:} he 
past three decades have witnessed an explosion in our abilities to control the behaviour of light, and in our understanding of the possible structures and topologies of electromagnetic radiation~\cite{torres-torner-twisted-photons-2011, gbur-singular-optics-2016, rubinsztein-structured-light-2017}. Building on the initial discoveries of wavefront dislocations and phase singularities~\cite{nye-dislocations-in-wave-trains-1974}, the field of singular optics now spans from optical communication technology~\cite{wang-otpical-communications-2012} through imaging~\cite{furhapter-spiral-imaging-2005}, the mechanical manipulation of matter~\cite{garces-volke-oam-transfer-2002,padgett-tweezers-with-a-twist-2011} and XUV/x-ray applications~\cite{hernandez-garcia-xuv-vortices-2017}, to a detailed understanding of the classical and quantum natures of the angular momentum of light~\cite{barnett-spin-orbital-2016, molina-twisted-photons-2007}.

Some of the most fascinating structures discovered by this programme are the topological features of light: recent work has described, sometimes experimentally, light fields with intricate knots in their field lines~\cite{kedia-knots-2013} and their optical vortices~\cite{leach-knotted-threads-of-darkness-2004, dennis-vortex-knots-2010, sugic-knotted-hopfion-2017}, as well as fields with spirals~\cite{freund-classification-cones-spirals-strips-2005}, umbilics~\cite{nye-wave-structure-1987, dennis-morphology-2002}, and Möbius strips~\cite{freund-optical-moebius-I-2010, freund-optical-moebius-II-2010, bauer-observation-mobius-2015, bauer-optical-mobius-2016, galvez-multitwist-2017, garcia-etxarri-moebius-2017} in their polarization. %
These structures are often associated with the invariance of the light field under coordinated rotations: that is, symmetry operations that rotate the spatial dependence of the fields by an angle $\theta$ and the fields' polarization by a multiple $\gamma \mspace{1mu} \theta$ of that angle, which means that they are associated with `mixed' angular momenta~\cite{ballantine-many-ways-to-spin-a-photon-2016} of the form $L_z + \gamma  \mspace{1.5mu} S_z$, where $L_z$ and $S_z$ are the orbital and spin angular momenta of light about the symmetry axis; these are conserved separately in the paraxial approximation, and can be measured independently~\cite{barnett-spin-orbital-2016, leach-interferometric-2004}.

For monochromatic light, the rotation coordination parameter $\gamma$ must be either an integer or a half-integer~\cite{nye-wave-structure-1987, ballantine-many-ways-to-spin-a-photon-2016}, since the only internal symmetry of a polarization ellipse is a rotation by $\pi$ over a half-period delay. However, general electromagnetic fields are not subject to this restriction: as a simple example, a three-fold rotational symmetry is possible by combining a circularly-polarized field with a counter-rotating second harmonic~\cite{kessler-lissajous-singularities-2003, freund-bichromatic-2003}, a configuration that forms a so-called `bicircular'~\cite{fleischer-spin-2014} trefoil-shaped Lissajous figure. The Lissajous singularities of bichromatic fields have been the object of some study~\cite{freund-polychromatic-singularities-2003, freund-polarization-critical-points-2003, yan-dynamical-lissajous-2010, chen-propagation-2017}, but their rotational properties have thus far largely gone unexplored.

In this work we use the higher internal symmetry of bicircular fields to construct and characterize beams of electromagnetic radiation that are invariant under coordinated rotations, as generated by $L_z + \gamma  \mspace{1.5mu} S_z$, for an arbitrary coordination parameter $\gamma$. Topologically, these beams' polarization corresponds to a torus knot, characterized by two indices: the order $n$ of internal symmetry of the polarization Lissajous figure, and the number of internal-symmetry rotations produced by a spatial traversal around the central singularity; for each internal symmetry, the latter forms a topologically-protected winding number of the electromagnetic field.

Moreover, we show how this winding number arises naturally as the phase winding number of the multipolar components of the third- and higher-order field tensor moments $\langle E_i E_j\cdots E_k\rangle$, in analogy to the monochromatic polarization ellipse's appearance as the quadrupole component of the polarization matrix $\langle E_i E_j \rangle$. Finally, we experimentally demonstrate these torus-knot beams and characterize their winding number via nonlinear polarization tomography.

\begin{figure*}[t]
\setlength{\fboxsep}{0pt}%
\setlength{\fboxrule}{1pt}%
\begin{tabular}{c}
  \subfloat{ \label{fig-initial-trefoil}
  \includegraphics[scale=1]{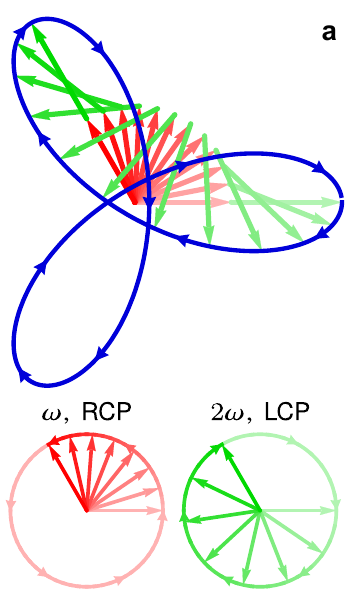} }
  \hspace{1mm}
  \subfloat{ \label{fig-initial-ferris-wheel}
  \includegraphics[scale=1]{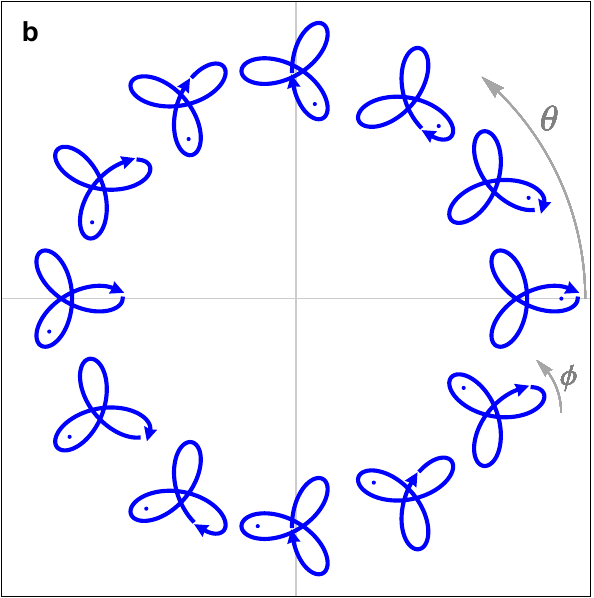} }
  \hspace{1mm}
  \subfloat{ \label{fig-buildup-helix}
  \includegraphics[scale=1]{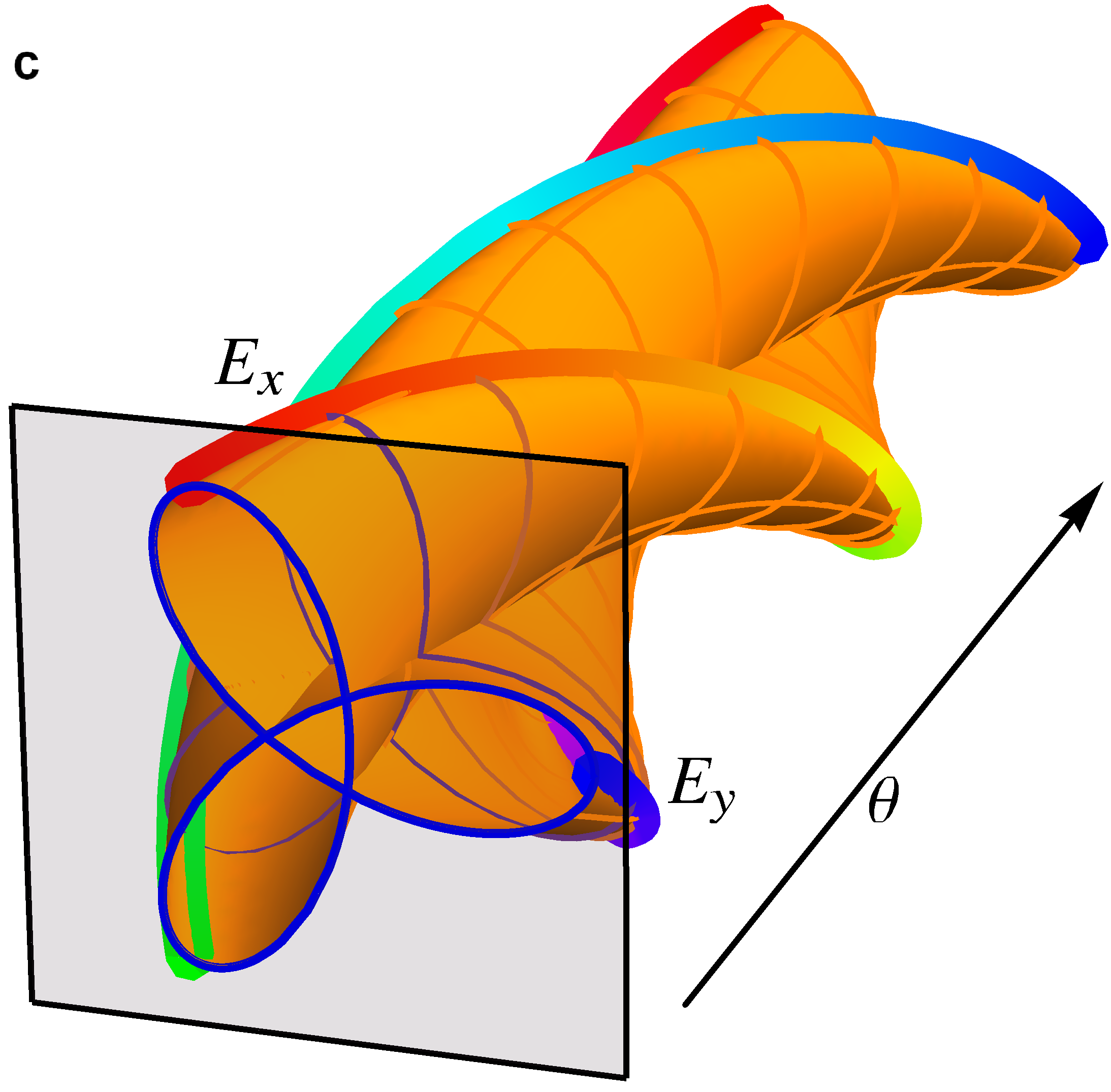} }
  \\[2mm]
  \subfloat{ \label{fig-twisted-trefoil-knot}%
  \includegraphics[scale=1]{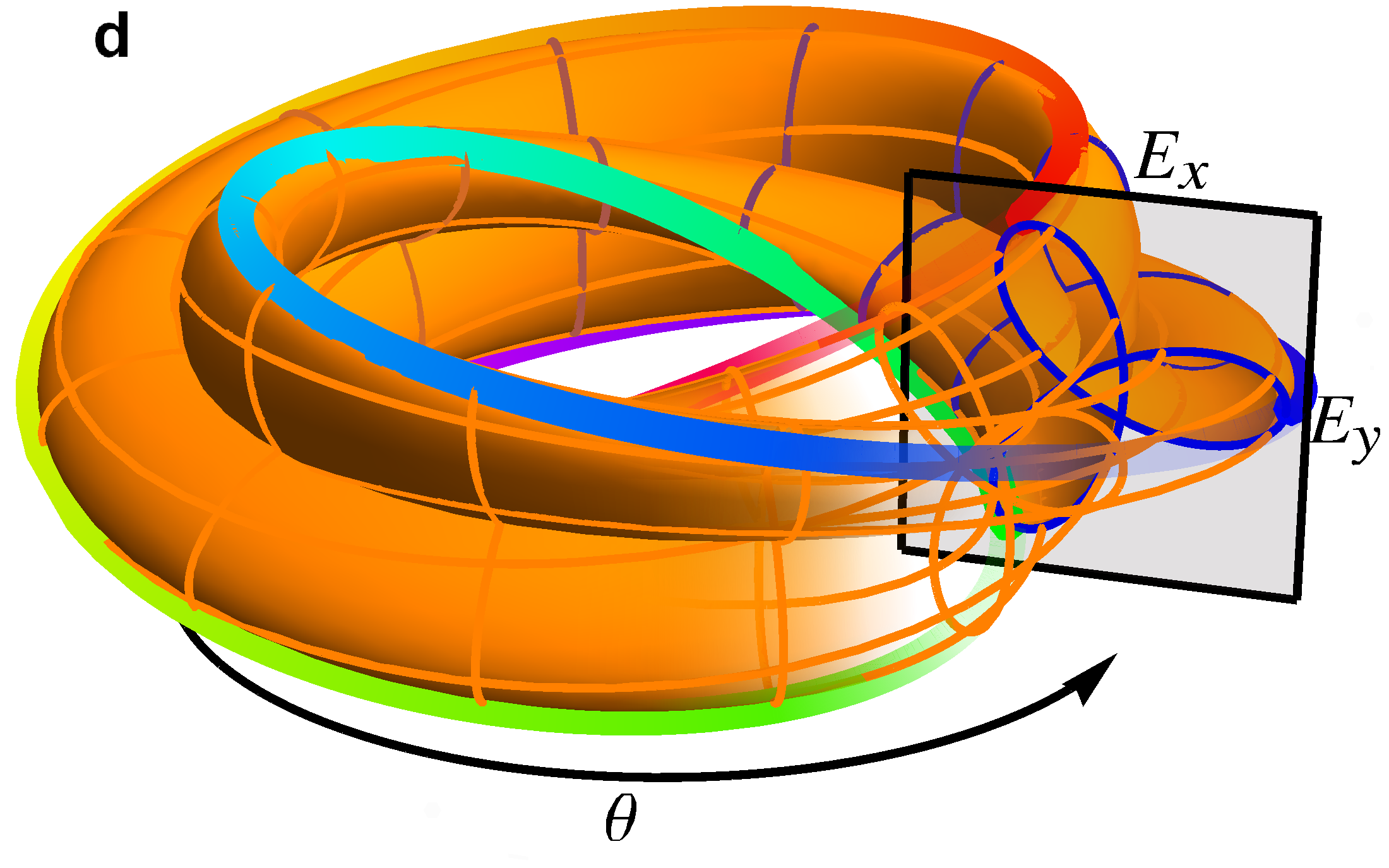} }
  \hspace{-1mm}
  \subfloat{ \label{fig-initial-knot}%
  \includegraphics[scale=1]{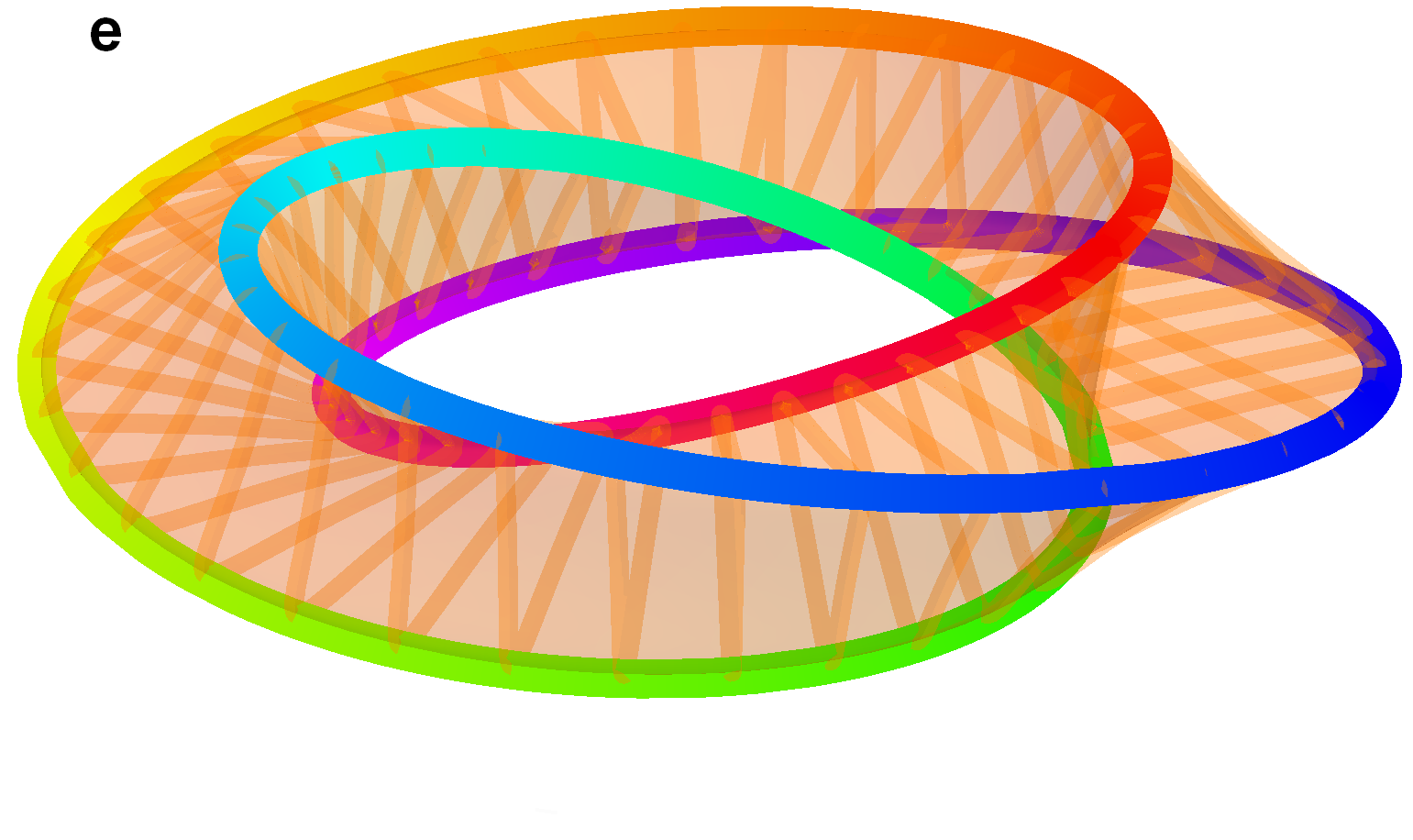} }
  \hspace{-1mm}
  \subfloat{ \label{fig-flat-torus-knot}%
  \includegraphics[scale=1]{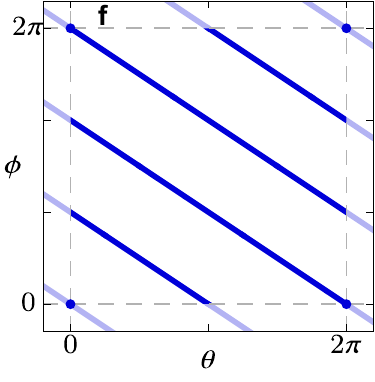} }%
\end{tabular}
\figurecaption{%
Coordinated-rotation invariance and torus-knot beam topology%
}{%
\textbf{(a)} The superposition of a right-circularly polarized beam (red) with its left-circularly polarized second harmonic (green) produces a trefoil-shaped polarization Lissajous figure (blue). The orientation of this trefoil depends on the relative phase between the two components, which can be made to vary along the azimuthal position $\theta$ by adding different orbital angular momenta $m_1$ and $m_2$ to the two components, shown in \textbf{(b)} for $m_1=0$ and $m_2=-2$: the field then has coordinated-rotation invariance of the form $R(\gamma\alpha)\vbe(\theta-\alpha, t)= \vbe(\theta,t+\tau\alpha)$ with coordination parameter $\gamma=-2/3$ and $\tau=1/6\omega$, and the lobe marked with a dot does not return to itself after a $2\pi$ azimuthal traversal over $\theta$. %
\textbf{(c)} To study this field's topology, we unfold the polarization and azimuthal dependence, and then \textbf{(d)} twist the resulting cylinder to reconnect the planes at $\theta=0$ and $2\pi$. %
\textbf{(e)} If we then retain only the paths of the tips of the trefoil lobes (colored by hue on \textbf{(c-e)} for visual clarity only), we obtain a knotted curve embedded on a torus, in this case the $(-2,3)$ torus knot. %
\textbf{(f)} This torus knot can be seen as the path of the lobes on the flat torus $[0,2\pi)\times[0,2\pi)$ of the azimuthal and polarization angles $\theta$ and $\phi$, but also as the coordinated rotations when seen as a subgroup of the independent-rotations group $\mathrm{SO}(2) \times \mathrm{SO}(2)$.
3D-printable models of \textbf{(d)} and \textbf{(e)} are available in \citer{supplementary-information}.
}
\label{fig-knot-buildup}
\end{figure*}

\section{Coordinated-rotation invariance}
For monochromatic light, rotational invariance is normally framed by requiring that the complex amplitude of the electric field obey an eigenvalue equation of the form
\begin{equation}
\hat{\mathcal{G}} \mspace{1.5mu} \tve(\vbr) = g \mspace{1.5mu} \tve(\vbr),
\label{invariance-as-eigenvalues}
\end{equation}
for some symmetry generator $\hat{\mathcal{G}}$. %
For polychromatic superpositions, however, there is no longer a single complex amplitude to which an eigenvalue equation like~\eqref{invariance-as-eigenvalues} can be applied -- and indeed, %
if one looks for invariant states by applying the generator $\hat{\mathcal{G}} = \hat L_z + \gamma  \mspace{1.5mu} \hat S_z$ separately to each amplitude and asking for joint eigenstates, one becomes (erroneously) restricted to the conclusions of the monochromatic case.

To deal effectively with the rotational invariance of polychromatic beams, then, it is crucial to realize that, even in the monochromatic case, the real-valued physical field $\vbe(\vbr,t) = \Re\mathopen{}\left[\tve(\vbr)e^{-i\omega t}\right]\mathclose{}$ is never an eigenstate of the symmetry generator. Instead, the rotational invariance of the physical force fields describes an equivalence between the symmetry operation and a time translation: for coordinated rotations, this reads
\begin{equation}
R(\gamma \alpha)\vbe \big(R^{-1}\mspace{-2mu}(\alpha)\vbr,t \big)
=
\vbe(\vbr, t+\tau \alpha)
,
\label{core-invariance-equation}
\end{equation}
where $R(\alpha)$ is a rotation matrix by angle $\alpha$ about the beam's symmetry axis, and $\tau$ is a constant with dimensions of time.

This invariance condition is fulfilled trivially by cir\-cu\-larly-polarized orbital-angular-momentum beams~\cite{barnett-spin-orbital-2016}, which are separately invariant under rotations of the image and the polarization. (Moreover, those rotations can be independently implemented when restricted to a single axis, using half-wave plates and Dove prisms, respectively~\cite{ballantine-many-ways-to-spin-a-photon-2016}.) Going beyond that, one can also form solutions of~\eqref{core-invariance-equation} which are invariant under the combined transformation but not under either of the separate ingredients: this is the case, for instance, for the `lemon' and `star' umbilic ellipse points~\cite{nye-wave-structure-1987, dennis-morphology-2002}, as well as the flat Möbius bands produced by conical refraction~\cite{ballantine-many-ways-to-spin-a-photon-2016}. These require an integer or half-integer~$\gamma$, since only rotations by multiples of $\pi$ will return a monochromatic polarization ellipse to itself, and they can generally be decomposed as superpositions of circularly-polarized beams of different orbital angular momentum.

To generalize these beams, we look to bichromatic combinations with higher-order internal rotational symmetry in the polarization, as provided by the bicircular combination~\cite{fleischer-spin-2014} shown in figure~\ref{fig-initial-trefoil}: we superpose a right-handed circularly polarized (RCP) beam at a fundamental frequency $\omega_1=\omega$ with its second harmonic at $\omega_2=2\omega$ on a left-circular polarization (LCP). Thus, over one-third of a period of the fundamental, the former rotates counter-clockwise by $\SI{120}{\degree}$, while the latter rotates clockwise by $\SI{240}{\degree}$, so the polarization combination is rotated rigidly by $2\pi/3$. More generally, combining counter-rotating beams at $p\omega$ and $q\omega$ will give a $(p+q)$-fold-symmetric Lissajous figure, but we focus on the $\omega$-$2\omega$ combination for simplicity.

\begin{figure}[t]
\centering
\includegraphics[width=0.49\textwidth]{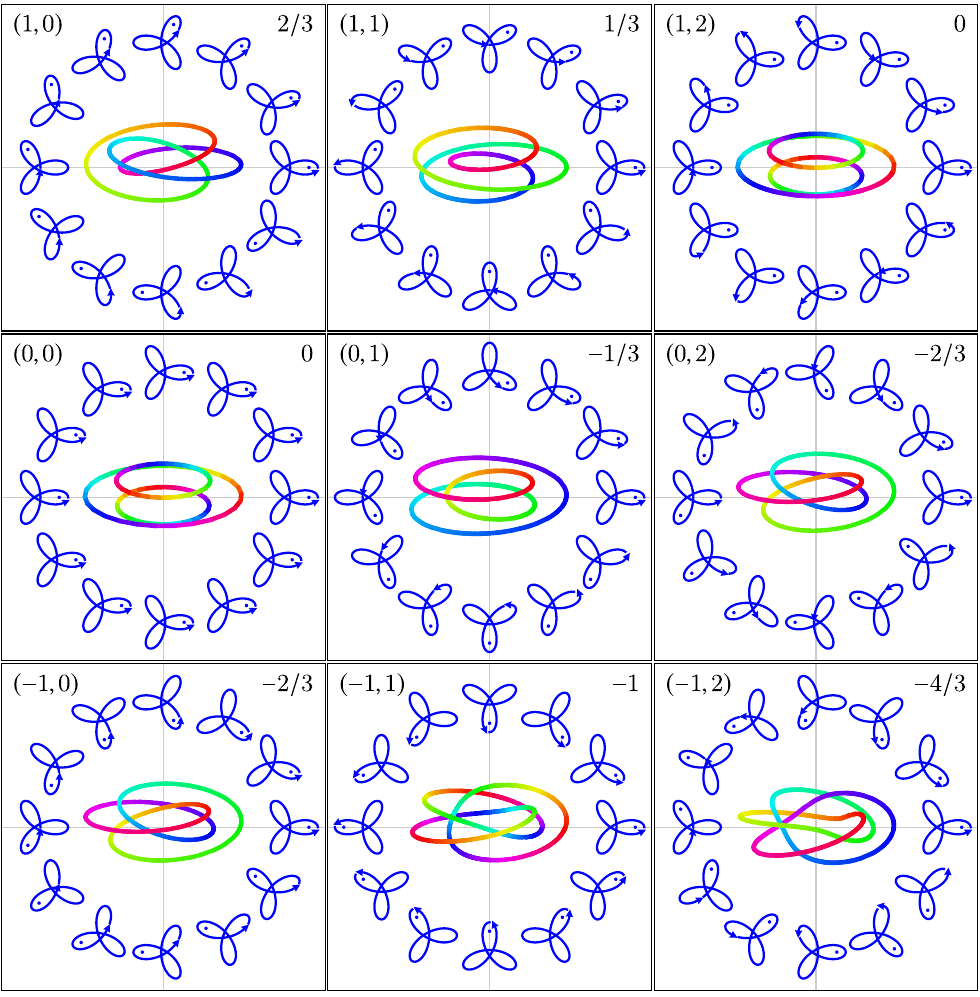}
\label{fig-knot-table}
\figurecaption{%
Possible topologies of torus-knot beams%
}{%
As the orbital angular momenta $(m_1,m_2)$ of the $\omega$ and $2\omega$ components is varied (upper left of each panel), the beam topology ranges over all $(m,3)$ torus knots, and the coordination parameter $\gamma$ (upper right of each panel) ranges over all multiples of $1/3$. When $m=m_2 p - m_1 q=2m_2-m_1$ is divisible by three, and $\gamma$ is an integer, the torus knot separates into three (possibly linked), distinct rings, reflecting the fact that at $\alpha=2\pi$ the coordinated rotation returns the trefoil lobes to their initial positions.
}
\end{figure}

Finally, to extend this symmetry to coordinated rotations, we work as in the monochromatic case and give the two light fields at frequencies $\omega_1$ and $\omega_2$ different orbital angular momenta $m_1$ and $m_2$, as exemplified in figure~\ref{fig-initial-ferris-wheel}: this causes the bicircular trefoil to take different orientations at different azimuthal positions around the beam, and a traversal by $2\pi$ around the axis produces a rotation by a fraction
\begin{equation}
\gamma
= 
\frac{
  m_2 \omega_1 - m_1 \omega_2
  }{
  \omega_1+\omega_2
  }
= 
\frac{
  m_2 p - m_1 q
  }{
  p+q
  }
\label{gamma-main-equation}
\end{equation}
of a revolution. For the $\omega$-$2\omega$ combination, with $p=1$ and $q=2$, this can be any arbitrary integer multiple of $1/3$; for arbitrary commensurate frequencies $\omega_1$ and $\omega_2$, $\gamma$ can be any rational number. (Irrational $\gamma$, on the other hand, are possible by using non-commesurate frequencies, though that requires a quasi-periodic field of infinite duration.)

Moreover, once the invariance under coordinated rotations has been formulated as in equation~\eqref{core-invariance-equation}, one can then solve for the most general field with that symmetry; we sketch the proof in the Methods section. When this general solution is restricted to only monochromatic contributions, one recovers the $\gamma \in \tfrac12 \mathbb Z$ restriction.

\section{Beam topology}
We have, then, beams with a trefoil polarization which rotates smoothly when moving around the axis of propagation of light in the plane perpendicular to this axis, coming back to the same trefoil but with a nontrivial internal rotation, as shown in the `Ferris-wheel' diagram in figure~\ref{fig-initial-ferris-wheel}. For monochromatic beams, this internal rotation induces a Möbius-strip topology to the polarization~\cite{freund-optical-moebius-I-2010, ballantine-many-ways-to-spin-a-photon-2016}. (This strip can also be lifted to a fully three-dimensional one if required~\cite{bauer-observation-mobius-2015, bauer-optical-mobius-2016}, but we restrict our attention to the topological Möbius strip in two dimensions.) In our case, the induced topology is different, but it can still be analyzed as in the planar monochromatic case, by following the tips of the trefoil over their orbit under the transformation.

In this spirit, then, we unfold the beam's polarization as shown in figure~\ref{fig-buildup-helix}, with the beam's trefoil polarization set against the azimuthal coordinate $\theta$, giving a corkscrew variation that terminates at a point, $\theta=2\pi$, identical to the initial $\theta=0$; to complete the visualization, we twist this corkscrew around to join these two equivalent points, as shown in figure~\ref{fig-twisted-trefoil-knot}. Here the key information is in the path followed by the tips of the trefoil, and the details of the path of the Lissajous figure can be distilled away as in figure~\ref{fig-initial-knot} to leave only the trefoil-tip path.

In this view it becomes clear that there is only one such path, which wraps around the figure three times before returning to its initial position; that is, all three lobe tips can be connected by the coordinated rotation. (However, this property disappears if $\gamma$ is an integer.) Moreover, in this view, it becomes clear that the lobe-tip path is a curve embedded on the surface of a torus, which immediately classes it as a torus knot~\cite{adams-knot-book-2004}, and in the example shown, the curve is indeed knotted: within this representation, it cannot be deformed smoothly to a simple unknotted loop.

Generally speaking, a $p\omega$-$q\omega$ bicircular combination with orbital angular momenta $m_1$ and $m_2$ on the two components will similarly have the lobe tips confined to a torus surface, and the resulting torus-knot path traced by the lobe tips can be characterized by two winding numbers:
\begin{itemize}[topsep=0pt,itemsep=-1ex,partopsep=1ex,parsep=1ex, label={\bfseries--}]
\item the number $n=p+q$ of times it passes any given fixed-$\theta$ cross section of the torus, equal to the number of lobes of the Lissajous figure, and
\item the (signed) number $m=m_2 p - m_1 q$ of times it crosses the inner diameter of the torus;
\end{itemize}
the knot is then labelled as the $(m,n)$ knot. If $m$ and $n$ admit a common divisor $d=\gcd(m,n)$, then the lobe-tip path separates into $d$ separate components, each of which crosses the inner diameter $m/d$ times; in that case, the components are $(m/d,n/d)$ torus knots pairwise linked with each other. The torus-knot order $(m,n)$, then, behaves almost exactly like a fraction, and indeed it is in direct correspondence with the rational coordination parameter $\gamma=m/n$. 

This includes, in particular, the ellipse-point umbilics of the monochromatic case, whose Möbius-strip topology~\cite{freund-optical-moebius-I-2010, ballantine-many-ways-to-spin-a-photon-2016} can be re-cast as the topology of the $(m,2)$ torus knot, which then admits an immediate generalization to other members of the knot family when the monochromatic restriction is broken.

\begin{figure*}[t!]
\newlength{\exptfigwidth}
\setlength{\exptfigwidth}{0.35\textwidth}
\setlength{\tabcolsep}{0.mm}
\newcommand{\figthreescale}{0.8}
\centering
\begin{tabular}{cc}
\begin{tabular}{c}
\subfloat{ \label{fig-experimental-setup-basic}
\includegraphics[width=0.45\textwidth]{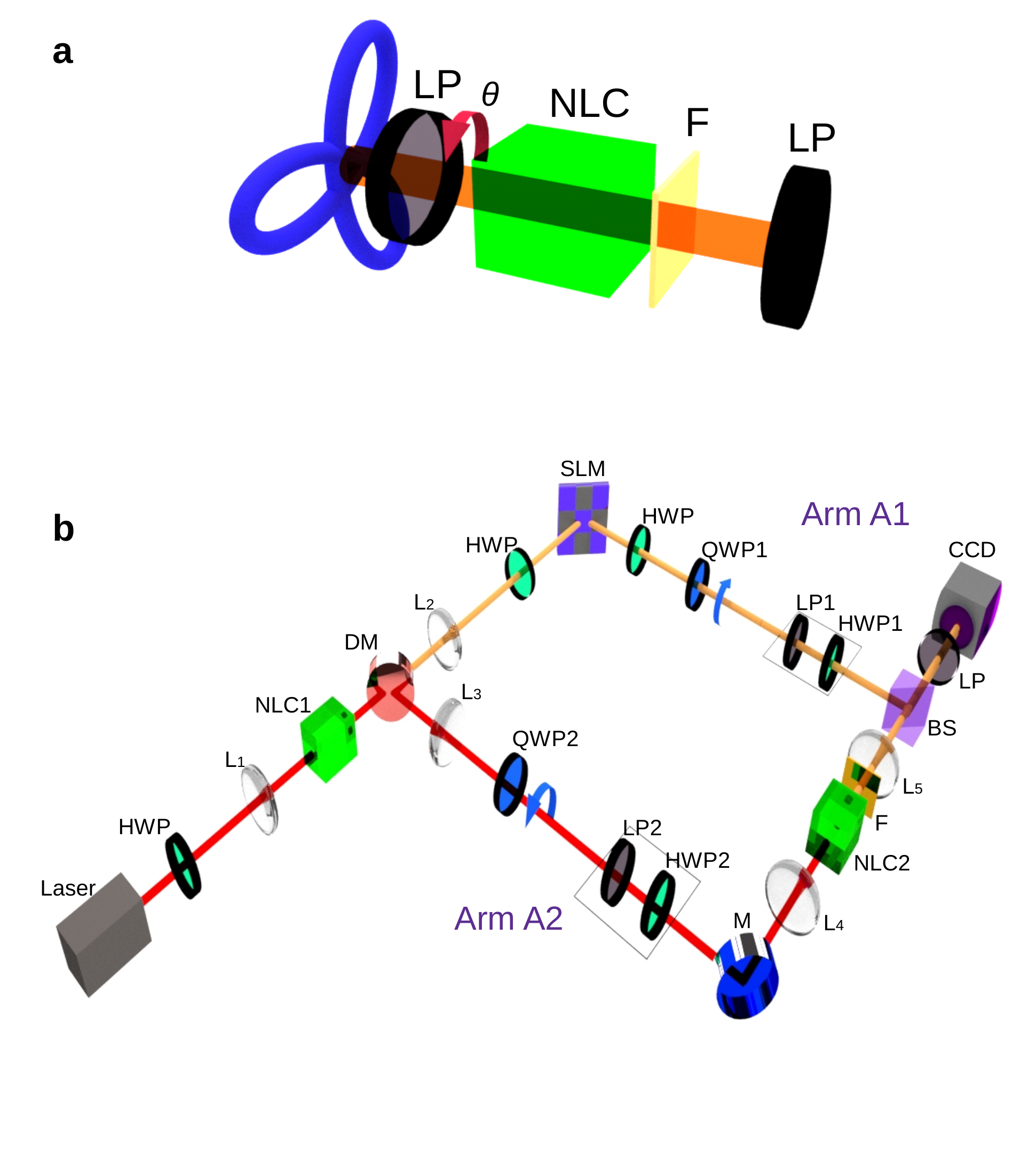}
}%
\subfloat{ \label{fig-experimental-setup}}%
\end{tabular}
&
\begin{tabular}{c}
{\sffamily \bfseries \small experiment} \\[-3.5mm]
\subfloat{ \label{fig-vortex-result-1}
\includegraphics[scale=\figthreescale]{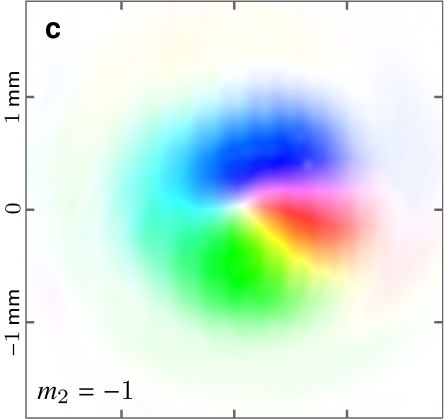}
} \\[-4mm]
\subfloat{ \label{fig-vortex-result-2}
\includegraphics[scale=\figthreescale]{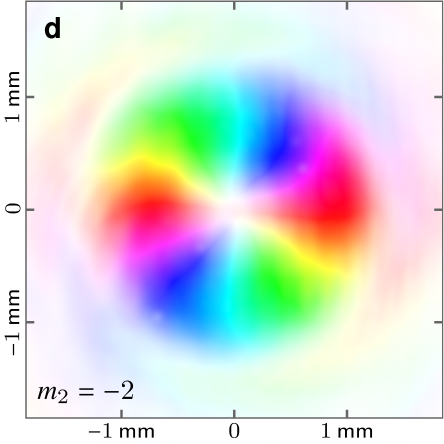}
} 
\end{tabular}
\hspace{-2.5mm}
\begin{tabular}{c}
{\sffamily \bfseries \small theory} \\[-3.5mm]
\subfloat{ \label{fig-polarization-field-1}
\includegraphics[scale=\figthreescale]{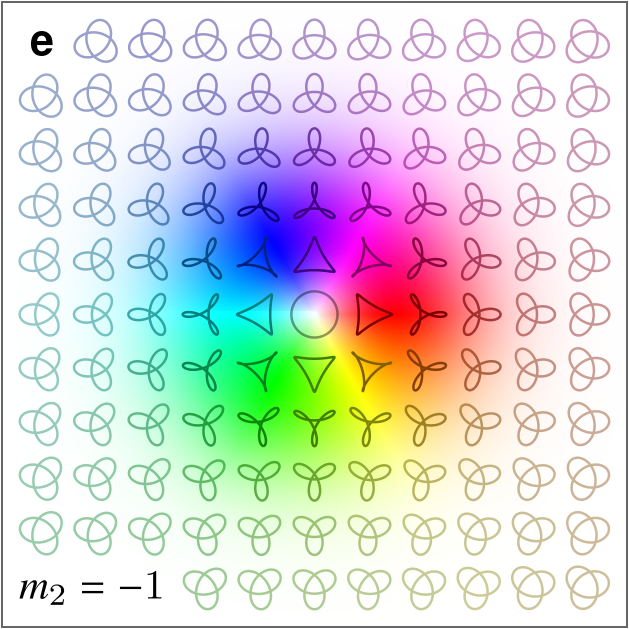}
} \\[-4mm]
\subfloat{ \label{fig-polarization-field-2}
\includegraphics[scale=\figthreescale]{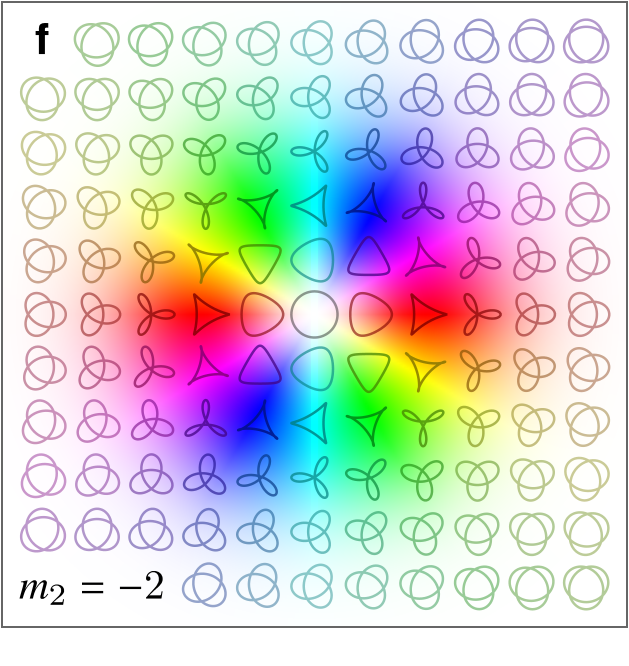}
}
\end{tabular}
\end{tabular}
\figurecaption{Experimental configuration and results}{
\textbf{(a)}~Basic scheme for nonlinear polarimetry, giving rise to %
\textbf{(b)}~our experimental set-up (see the Methods section for a detailed description and definitions of the acronyms).
\textbf{(c,d)}~Experimental measurement of $T_{3,3}(x,y)$ for $m_2=-1$ and $m=-2$ on the $2\omega$ arm, taken by Fourier-transforming the CCD images as a function of the polarizer angles, shown with its amplitude $|T_{3,3}(x,y)|$ as the saturation and the phase $\mathsf{arg}(T_{3,3}(x,y))$, directly linked to the trefoil orientation angle, as the hue. 
\textbf{(e,f)}~Theoretical prediction for the field tensor moment $T_{3,3}(x,y)$ of a gaussian and a Laguerre-Gauss beam with $m_2=-1$ and $m=-2$, showing a clear match to the corresponding measurements, overlaid with the Lissajous figures of the polarization at different points around the beam. 
At the center of the beam the Lissajous figure is a circle, and it shifts to a triangle and then a trefoil as the probe point gets further from the axis -- but the orientation of the deformation away from circularity depends on the direction of departure from the axis. 
}
\label{fig-setup-and-results}
\end{figure*}

The torus-knot topology of the beam also has a natural algebraic interpretation, as the symmetry subgroup of transformations, generated by $J_\gamma = \hat{L}+\gamma\hat{S}$, formed by the coordinated rotations. The decoupling of spatial and polarization rotations as independent symmetries of the field unfolds the $\mathrm{SO}(2)$ invariance of the paraxial theory into the direct product of two copies of that group, $\mathrm{SO}(2) \times \mathrm{SO}(2)$, which is naturally identified as the flat torus. The coordinated rotations, for a given $\gamma$, form a one-dimensional continuous closed subgroup of that torus, which winds $(m,n)$ times about the torus's two dimensions: in other words, it is precisely the same torus knot as the beam topology, and as shown in \reffig{fig-flat-torus-knot}. Because of this algebraic identification, we refer to the subgroup generator $J_\gamma$ as the torus-knot angular momentum (TKAM); with the expanded range of $\gamma$, this now forms a natural alternative to fractional-OAM beams~\cite{gotte-fractional-oam-2008}.

Torus knots have appeared several times in optical contexts, and it is possible to form such knots using both optical vortices~\cite{leach-knotted-threads-of-darkness-2004, dennis-vortex-knots-2010, sugic-knotted-hopfion-2017} and field lines~\cite{kedia-knots-2013}, as explicit three-dimensional objects. Here, on the other hand, the torus knot does not exist as a real-space three-dimensional object; it is, instead, a characterization of the beam's topology as a planar Lissajous-figure field, generalizing the Möbius-strip topology of monochromatic ellipse points.

\section{Orientation measures and their singularities}
At its heart, the optical singularity at the beam axis of a coordinated-rotation-invariant beam is a singularity in the orientation of the polarization Lissajous figure, as exemplified in Figs.~\ref{fig-polarization-field-1} and~\ref{fig-polarization-field-2}. As such, to fully characterize it, we need a numerical measure of this orientation. 
The existing orientation measures~\cite{freund-coherency-matrix-2004, freund-bichromatic-2003, freund-polychromatic-singularities-2003, freund-polarization-critical-points-2003} are based on the polarization matrix $\langle E_i E_j\rangle$ of the beam, but this matrix is inappropriate to the fields studied here because it is invariant under $\SI{180}{\degree}$ rotations, and it is therefore blind to structures with three-fold or higher symmetries.

The naive extension of this approach is to change the time-averaged product of two field components for three of them: that is, the rank-three field tensor moment $\langle E_i E_j E_k\rangle$. This object is somewhat too large to analyze directly, at four independent components, but since we are looking for its transformation properties under rotations, it is natural to decompose it into the two representations it carries, $\ell=3$ and $\ell=1$, of the planar rotation group $\mathrm{SO}(2)$. Thus, the orientation of the bicircular Lissajous figure is best characterized through the multipole components
\begin{subequations} %
\begin{align}
T_{3,3} 
& =
\left\langle \left(E_x(t)+iE_y(t)\right)^3 \right\rangle
\ \text{and} 
\label{t33-definition}
\\
T_{1,3}
& = 
\left\langle \left(E_x(t)+iE_y(t)\right)\left(E_x(t)^2+E_y(t)^2\right) \right\rangle,
\label{t13-definition}
\end{align}
\label{t13-t33-definition}%
\end{subequations}%
with $\alpha_3=\frac{1}{3}\arg(T_{3,3}) \mod \SI{120}{\degree}$, the phase of the hexapole component, giving the trefoil orientation angle. Further, the count of how many times $\alpha_3$ loops about its range over a spatial traversal around the beam axis provides a new to\-po\-lo\-gi\-cally-protected winding number in direct correspondence with $\gamma$.

Here the presence of the dipole representation in \eqref{t13-definition} is an added bonus: it provides a separate winding number, and it points to the presence of `true' vector vortices (which cannot split into a pair of ellipse points), with a singularity in the dipole orientation angle $\alpha_1=\arg(T_{1,3})$. This can be achieved, for example, with co-rotating cir\-cu\-lar\-ly-polarized fields at frequencies $\omega$ and $2\omega$, which produces a cardioid-shaped Lissajous figure with a clear directionality.

For both field tensors, the cubic nature of the polynomial inside the time-average places strict restrictions on the field combinations that can contribute. Thus, for a bichromatic superposition of the form $\vbe(t) = \sum_\pm\Re\big[E_{1,\pm} \ue{\pm}e^{-i\omega t}+E_{2,\pm} \ue{\pm} e^{-2i\omega t} \big]$ with $\ue\pm = \tfrac{1}{\sqrt{2}}(\ue{x}\pm i\ue{y})$, the time average covers several products of exponentials, but only the ones at zero total frequency are retained: that is, products of the form $E_1^2 E_2^*$, or, more precisely,
\begin{align}
T_{3,3}
& =
\frac{3}{\sqrt{8}}\bigg[ E_{1,{-}}^{\: 2} E_{2,{+}}^* + {{E_{1,{+}}^{*}}^{\!\!\!\!2}} \, E_{2,{-}}^{\phantom{*}} \bigg]
\label{t33-explicit}
\end{align}
for the hexapole tensor component, with the $E_1^2 E_2^*$ dependence mirroring the nonlinear processes that must be used to phase-lock the two components of the field, and which we use below to measure the trefoil orientation.

On a more general setting, the definitions in \eqref{t13-t33-definition} generalize transparently to characterize $p\omega$-$q\omega$ combinations via the $\ell$-polar component of the field tensor of rank $n=p+q$,
\begin{align}
T_{\ell,n}
& = 
\left\langle \left(E_x(t)+iE_y(t)\right)^\ell \left(E_x(t)^2+E_y(t)^2\right)^\frac{n-\ell}{2} \right\rangle.
\label{tlk-definition}
\end{align}
This reproduces, via $T_{2,2}$, the existing ellipse orientation measures for the monochromatic case, and it extends to a bi-infinite family of optical topological winding numbers -- which is relevant even in the $\omega$-$2\omega$ case, where $T_{2,4}$, $T_{2,6}$ and even $T_{2,8}$ are required, on dimension-counting grounds, to fully characterize the quadrupolar orientation of general Lissajous figures. Similarly, if the polarization is taken out of the paraxial regime, the $\mathrm{SO}(3)$ multipole components of the field tensor are the natural arena to describe the shape of the fields and its possible singularities.

\section{Experimental observation}
The generation of bicircular $\omega$-$2\omega$ beams with coordinated-rotation invariance is relatively simple, but their detection poses additional challenges, since linear-optical polarimetry is insensitive to the relative phase between the two chromatic components (unless the entire beam is transformed by a physical coordinated rotation~\cite{leach-interferometric-2004}), and therefore to the bicircular trefoil's orientation. To measure this phase, then, we require a nonlinear polarization tomography, with a quadratic second-harmonic-generation step that echoes the beam generation step and up-converts the phase of the fundamental so it can be compared with the second harmonic's.

For monochromatic fields, the orientation of the polarization ellipse is normally measured by inserting a linear polarizer and rotating its direction $\theta$ from $0$ through $2\pi$; the orientation angle can then be extracted as the phase of the second Fourier component of the measured intensity $I(\theta)$, mirroring its appearance as the argument of the second field tensor $T_{2,2}$ (itself a re-expression of the polarization matrix $\langle E_iE_j\rangle$). Similarly, for bicircular trefoils, whose orientation is encoded in $T_{3,3}$, we look for a polarimeter that will encode the bichromatic polarization as the $3\theta$ component of the intensity as the polarimeter is rotated.

In an idealized sense, this can be achieved by projecting both colours' polarizations on a linear polarizer oriented at angle $\theta$, followed by a type~0 second-harmonic generation step on the same axis and a filter to block the fundamental, as shown in \reffig{fig-experimental-setup-basic}, which then feeds to an imaging system. The interference between the initial and the detection-stage $2\omega$ light then produces a $3\theta$ component proportional to $T_{3,3}$ -- and, in addition, the first Fourier component is proportional to $T_{1,3}$. In practice, physical rotations of the quadratic crystal make for challenging alignment, so we use an optically-equivalent system sketched in \reffig{fig-experimental-setup} and described in the Methods section.

The measured Fourier components, shown in Figs.~\ref{fig-vortex-result-1} and \ref{fig-vortex-result-2}, clearly exhibit a nonzero winding number, in agreement with the theoretical predictions shown in Figs.~\ref{fig-polarization-field-1} and~\ref{fig-polarization-field-2}. This maps directly into a nonzero winding number of the third field moment tensor $T_{3,3}$, and it acts as a smoking-gun sign of coordinated-rotation invariance with a coordination parameter $\gamma$ outside the half-integral constraints of the monochromatic case.

\section{Outlook}
Our results provide a set of topologies that are achievable at optical polarization singularities when the monochromatic restriction is lifted: beams with well-defined torus-knot angular momentum $J_\gamma= L+\gamma S$, invariant under the coordinated rotations generated by $J_\gamma$, are possible for any rational mixing parameter $\gamma=m/n$, producing a polarization field in the shape of the $(m,n)$ torus knot. We also present the theoretical and experimental toolset, via field moment tensors and their presence in nonlinear polarization tomography, that can be used to characterize these beams and their associated singularities and winding numbers.

These features can be used as building blocks for more elaborate field topologies, 
from non-paraxial equivalents that can exhibit polarization torus knots as explicit three-di\-men\-sional objects as in the Möbius-band case~\cite{bauer-observation-mobius-2015, bauer-optical-mobius-2016}, 
to knots of coordinated-rotation-invariant vortex cores, 
or the trefoils' equivalent to streamlines; 
they can also be extended to evanescent light and combined with complex light shaping of the longitudinal polarization component~\cite{maucher-polarization-structures-2018}. 
The association with fractional values of an angular momentum also opens the door to the simulation of anyonic behavior~\cite{wilczek-fractional-spin-1982, wilczek-fractional-statistics-1984} 
by using light's spin-orbit interaction with matter~\cite{cardano-spin-orbit-photonics-2015, bliokh-spin-orbit-2015, bliokh-geometrodynamics-2009} 
and Bose-Einstein condensates~\cite{lin-spin-orbit-bec-2011}, 
as well as imprinting the beam's torus-knot topology onto the state of an atomic condensate~\cite{maucher-knotted-bec-vortex-2016}. 
Similarly, since symmetry generators are generally conserved in nonlinear optics~\cite{bloembergen-conservation-1980}, the same should be true for $J_\gamma$ in both the perturbative and non-perturbative~\cite{fleischer-spin-2014, hickstein-non-collinear-2015} domains.

From a quantum electrodynamical perspective, the existence of polychromatic beams invariant under coordinated rotations challenges the view~\cite{ballantine-many-ways-to-spin-a-photon-2016} that invariant beams should be defined strictly as eigenstates of the torus-knot angular momentum generator $\hat{L}+\gamma\hat{S}$: the direct analogue to~\eqref{core-invariance-equation} is a condition of the form
\begin{align}
\begin{aligned}
e^{-i\alpha(\hat{L}_z+\gamma \hat{S}_z)/\hbar} |\psi\rangle
& =
e^{-i\tau\alpha\hat{H}/\hbar} |\psi\rangle
,\ \text{or equivalently}\\
(\hat{L}_z+\gamma \hat{S}_z)|\psi\rangle
& =
\tau\hat{H} |\psi\rangle
,
\end{aligned}
\label{quantum-invariance-equation}%
\end{align}
in its infinitesimal version. Nontrivial solutions of \eqref{quantum-invariance-equation} do exist, using different numbers of photons on each component, and they have a clean relationship with the operator version of~\eqref{t13-t33-definition}. However, the extent to which they can be extended to a complete basis of states requires further attention, as is the degree to which the TKAM $\hat{L}_z+\gamma \hat{S}_z$ can be considered a `true' angular momentum operator~\cite{barnett-spin-orbital-2016, ballantine-many-ways-to-spin-a-photon-2016} when the mixing parameter $\gamma$ steps out of the half-integer domain.

\section{Acknowledgements}
We thank Maria Maffei and Isaac Freund for helpful conversations, and Xavier Menino for 3D-printing assistance.
E.P. acknowledges Cellex-ICFO-MPQ fellowship funding. 
E.P., M.L. and A.C. acknowledge funding from the Spanish Ministry MINECO (National Plan 15 Grant: FISICATEAMO No. FIS2016-79508-P, SEVERO OCHOA No. SEV-2015-0522, FPI), European Social Fund, Fundació Cellex, Generalitat de Catalunya (AGAUR Grant No. 2017 SGR 1341 and CERCA/Program), ERC AdG OSYRIS, EU FETPRO QUIC, and the National Science Centre, Poland-Symfonia Grant No. 2016/20/W/ST4/00314.
V.V.H. gratefully acknowledges financial support from Secretaría de Ciencia, Tecnología e Innovación de la Ciudad de México.
J.P.T. acknowledges support from Generalitat de Catalunya (Program ICREA Academia). G.J.M. was supported by the Secretaria d'Universitats i Recerca del Departament d'Economia i Coneixement de la Generalitat de Catalunya, as well as the European Social Fund~-- FEDER.
A.P. acknowledges funding from Comunidad de Madrid through TALENTO grant ref. 2017-T1/IND-5432.
A.C. acknowledges financial support from the ERC Synergy Grant UQUAM, the SFB FoQuS (FWF Project No. F4016-N23), the UAB Talent Research program and from the Spanish Ministry of Economy and Competitiveness under Contract No.\ FIS2017-86530-P.

\section{Author contributions}
E.P. conceived the project and developed the theory. 
G.J., V.V.H, E.P. and J.P.T. designed the experiment. 
G.J. and V.V.H. conducted the experiment. 
A.P., A.C. and M.L. assisted with the theory. 
E.P. wrote the manuscript, with assistance from V.V.H. on the Methods section.
J.P.T. supervised the experimental work; 
M.L. oversaw the theory development.
All authors contributed to the scientific discussion.

\bibliographystyle{arthur} 
\bibliography{references}{}

\section{Methods}

\subsection{General solution to the invariance equation}
In this section we show the general solution to the field invariance property \eqref{core-invariance-equation} under coordinated rotations,
\begin{equation}
R(\gamma \alpha)\vbe \big(R^{-1}\mspace{-2mu}(\alpha)\vbr,t \big)
=
\vbe(\vbr, t+\tau \alpha)
,
\tag{\ref{core-invariance-equation}}
\end{equation}
where the rotation matrix acts as
\begin{equation}
R(\alpha)\vb v
=
\begin{pmatrix}
\cos(\alpha) & -\sin(\alpha) & 0 \\
\sin(\alpha) &  \,\,\cos(\alpha) & 0 \\
0 & 0 & 1
\end{pmatrix}
\begin{pmatrix}
v_x \\ v_y \\ v_z
\end{pmatrix}
,
\end{equation}
so its action as a passive transformation in cylindrical coordinates is $ \vbe(R(\alpha)^{-1} \vbr , t) = \vbe(r, \theta - \alpha, z, t)$. With that in hand, we can turn the invariance property \eqref{core-invariance-equation} into its local version by differentiating with respect to $\alpha$, which produces
\begin{subequations}
\begin{empheq}[left=\empheqlbrace]{align}
-\frac{\partial E_x}{\partial\theta} - \gamma E_y 
& = \tau \frac{\partial E_x}{\partial t}
,
\\
-\frac{\partial E_y}{\partial\theta} + \gamma E_x 
& = \tau \frac{\partial E_y}{\partial t}
.
\end{empheq}%
\label{moebius-condition-differential}%
\end{subequations}%
This is a pair of coupled partial differential equations, but since the derivatives appear with the same sign in both, we can reduce them to a single unified form by setting $u=\theta +t/\tau$ and $v=\theta-t/\tau$, so that they combine to a single complex equation,
\begin{equation}
\left(
-\frac{\partial}{\partial u} +\frac12 i\gamma
\right)
(E_x+iE_y)=0
,
\end{equation}
where the conjugate coordinate $v$ drops out, leading to the simple solution
\begin{subequations}%
\begin{align}
E_x+iE_y
& =
F(v)e^{\frac{i}{2}\gamma u}
\\ & =
F(\theta-t/\tau)e^{\frac{1}{2}i\gamma(\theta+t/\tau)}
\end{align}
\label{mid-step-general-solution}
\end{subequations}%
in terms of an arbitrary function $F(v) = F(\theta-t/\tau)$.

The solution in~\eqref{mid-step-general-solution}, however, is not quite complete, because of the possible fractional exponent in $e^{\frac{1}{2}i\gamma(\theta+t/\tau)}$, in a function that needs to be periodic in $\theta$, and this needs to be offset by setting $F(v) = \tilde F(v) e^{-\frac12 i \gamma v}$ for a periodic $\tilde F(v)$. This then requires that the solution be of the form
\begin{subequations}%
\begin{align}
E_x+iE_y
& =
\tilde F(\theta-t/\tau) 
e^{i\gamma t/\tau}
\\ & =
\sum_{m=-\infty}^\infty
F_m
e^{im\theta}
e^{i\frac{\gamma-m}{\tau}t}
,
\end{align}
\label{general-solution}
\end{subequations}%
in terms of the Fourier coefficients of $\tilde F(v)$. This completely fixes the possible dependence of $\mathbf E$, as a superposition of a discrete set of orbital angular momentum (OAM) modes $e^{im\theta}$ at prescribed frequencies $\left| \frac{\gamma-m}{\tau} \right|$, with right- (resp. left-)handed circular polarizations at negative (resp. positive) frequencies.

In particular, the general solution in~\eqref{general-solution} is enough to reproduce the restrictions of the monochromatic case, which only allows for two terms $m_1$ and $m_2$ to contribute to the sum, at positive and negative frequencies $e^{-i\omega t}$ and $e^{+i\omega t}$, which then requires that
\begin{equation}
\frac{\gamma -m_1}{\tau}
=
\omega
=
-\frac{\gamma - m_2}{\tau}
,
\end{equation}
and therefore restricts the coordination parameter
\begin{equation}
\gamma = \frac{m_1 + m_2}{2}
\end{equation}
to integer or half-integer values, as found in previous work.

On the other hand, if the general solution~\eqref{general-solution} is required to include right- and left-handed components with OAM $m_1$ and $m_2$ at frequencies $\omega_1$ and $\omega_2$, then $\gamma$ and $\tau$ are required to obey $\gamma + m_1  = \omega_1\tau$ and $\gamma - m_2  = - \omega_2\tau$, resulting in the arbitrary coordination parameter $\gamma$ given in \eqref{gamma-main-equation}, together with the delay constant $\tau = \frac{m_1 + m_2}{\omega_1+\omega_2}$.

\subsection{Nonlinear polarization tomography}
In this section we present the core mechanism for nonlinear polarization tomography of bichromatic $\omega$-$2\omega$ combinations, as schematized in \reffig{fig-experimental-setup-basic}; we consider its effect on an arbitrary bichromatic combination, but we neglect the spatial dependence for now. Thus, we consider an electric field of the form
\begin{equation}
\vbe(t)
=
\Re\mathopen{}\left[
\vbe_1 e^{-i\omega t} + \vbe_2 e^{-2i\omega t}
\right]\mathclose{}
,
\end{equation}
which then passes through
\begin{enumerate}[topsep=0.4em,itemsep=-1ex,partopsep=1ex,parsep=1ex, leftmargin=*]
\item a linear polarizer along $\uu= \cos(\theta)\ue{x} + \sin(\theta)\ue{y}$, which transforms the field to
\begin{equation}
\mathrm{LP}_\theta \vbe(t)
=
\uu
\Re\mathopen{}\left[
\uu\cdot\vbe_1 e^{-i\omega t} + \uu\cdot\vbe_2 e^{-2i\omega t}
\right]\mathclose{};
\end{equation}
followed by

\item a nonlinear crystal that produces type~0 second-harmonic generation along $\uu$, thereby adding a contribution
\begin{equation}
\vbe_\mathrm{SHG}(t)
=
\uu
\Re\mathopen{}\left[
\chi\left(\uu\cdot\vbe_1\right)^2 e^{-2i\omega t}
\right]\mathclose{}
\end{equation}
to the field, where $\chi$ is a combination of the crystal's quadratic susceptibility and the interaction length; and finally

\item a filter that eliminates the fundamental before the intensity is measured.
\end{enumerate}

Our core observable is therefore the measured intensity as a function of the rotation angle $\theta$, given by the time average
\begin{align}
I(\theta)
& =
\left<
\Re\mathopen{}\left[
\chi\left(\uu\cdot\vbe_1\right)^2 e^{-2i\omega t} + \uu\cdot\vbe_2 e^{-2i\omega t}
\right]\mathclose{}^2
\right>
\nonumber \\ & =
\frac12
\left[
( \chi\left(\uu\cdot\vbe_1^*\right)^2 + \uu\cdot\vbe_2^* ) 
( \chi\left(\uu\cdot\vbe_1\right)^2 + \uu\cdot\vbe_2 )
\right]
\! . \!
\end{align}
Writing the electric field in circular components as $\vbe_j=\sum_\pm E_{j,\pm} \ue{\pm}$ with $\ue\pm = \tfrac{1}{\sqrt{2}}(\ue{x}\pm i\ue{y})$ and $j=1,2$, as in the main text, so that the components along $\uu$ now read $\uu \cdot \vbe_j =  \tfrac{1}{\sqrt{2}} \left( E_{j,+}e^{i\theta} + E_{j,-}e^{-i\theta} \right)$, we can express the intensity as an explicit Fourier series  of the form
\begin{align}
I(\theta) 
& = 
I_0 + I_1(\theta) + I_2(\theta) + I_3(\theta) + I_4(\theta)
.
\end{align}
Here the even components, $I_0$, $I_2(\theta)$ and $I_4(\theta)$, carry limited information, as they contain no cross terms between $\vbe_1$ and~$\vbe_2$, and they are therefore insensitive to the coherence between the two components:
\begin{subequations}
\begin{align}
I_0
& = 
\frac{1}{8}\chi^2 \left( \left|\vbe_1\right|^4 + 2 \left|E_{1,-}E_{1,+}\right|^2 \right)
+ \frac14 \left|\vbe_2\right|^2 
,
\\
I_2(\theta)
& =
\Re\mathopen{}\left[
\left( \frac12 \chi^2|\vbe_1|^2 E_{1,-}^*E_{1,+} +  \frac12 E_{2,-}^*E_{2,+} \right) e^{2i\theta}
\right] \mathclose{}
, 
\\
I_4(\theta)
& =
\Re\mathopen{}\left[ \frac{1}{4}\chi^2 {{E_{1,{-}}^{*}}^{\!\!\!\!2}} E_{1,{+}}^{\: 2}e^{4i\theta} \right]
.
\end{align}
The odd components, $I_1(\theta)$ and $I_3(\theta)$, on the other hand, do contain the necessary $\vbe_1$-$\vbe_2$ coherences: they are given by
\begin{align}
I_1(\theta)
& =
\frac{\chi}{\sqrt{8}}
\Re\mathopen{}\left[
\left(
E_{1,+}^* \left( \vbe_1^*\cdot\vbe_2 + E_{1,-}^*E_{2,-} \right)
\right. \right. \\ \nonumber & \qquad \qquad \left. \left.
+ E_{1,-} \left( \vbe_1\cdot\vbe_2^* + E_{1,+}E_{2,+}^* \right)
\right)^*e^{i\theta} 
\right]\mathclose{}
\\
I_3(\theta)
& =
\frac{\chi}{\sqrt{8}} 
\Re\mathopen{}\left[
\left( E_{1,{-}}^{\: 2} E_{2,{+}}^* + {{E_{1,{+}}^{*}}^{\!\!\!\!2}} \, E_{2,{-}}^{\phantom{*}} \right)^* e^{3i\theta} 
\right]\mathclose{}
\end{align}
\end{subequations}
and, in fact, they are direct measures of the dipole and hexapole components of the third field moment tensor, as defined in \eqref{t13-t33-definition}, which are given in this context by
\begin{subequations}
\begin{align}
T_{3,3}
& =
\frac{3}{\sqrt{8}}\bigg[ E_{1,{-}}^{\: 2} E_{2,{+}}^* + {{E_{1,{+}}^{*}}^{\!\!\!\!2}} \, E_{2,{-}}^{\phantom{*}} \bigg]
\quad \text{and}
\\
T_{1,3}
& =
\frac{1}{\sqrt{8}}\left[
E_{1,{+}}^* \left( \vbe_1^* \cdot \vbe_2 + E_{1,{-}}^* E_{2,{-}}^{\phantom{*}}\right)
\right. \\ & \qquad \qquad + \left. \nonumber
E_{1,{-}}^{\phantom{*}} \left( \vbe_1 \cdot \vbe_2^* + E_{1,{+}}^{\phantom{*}} E_{2,{+}}^*\right)
\right]
.
\end{align}
\end{subequations}

\setcounter{figure}{0}
\renewcommand\thefigure{S\arabic{figure}}

\subsection{Experimental implementation}
We now describe our experimental implementation of the abstract nonlinear polarimetry delineated above, embodied in the Mach-Zehnder interferometer depicted in \reffig{fig-experimental-setup}.

We used as the pump source a Gaussian beam from a diode laser amplified with an erbium-doped fiber amplifier (EDFA), centered at $\SI{1550}{nm}$ with a power of $\SI{300}{mW}$, a beam diameter of $\SI{2.6}{mm}$ and vertical polarization, which is oriented using a half wave plate (HWP). 
The pump beam was sent through a lens $L_{1}$, with a focal length of $\SI{15}{cm}$, pumping a 10-mm-long periodically-poled lithium niobate (PPLN) nonlinear crystal (NLC1) which was placed at the focal distance of $L_{1}$ to generate second harmonic type 0 centered at $\SI{775}{nm}$. A dichroic mirror (DC) was placed after the NLC1 which transmits the light centered at $\SI{775}{nm}$ from the second-harmonic generation and reflects the light centered at $\SI{1550}{nm}$  from the pump, forming the two arms of the Mach-Zehnder interferometer.

The beam with wavelength centered at $\SI{775}{nm}$ was propagated trough the arm A1 and transmitted through the lens $L_{2}$, with focal length of $f = \SI{20}{cm}$, forming a telescope with $F_{1}$ to magnify the beam by a factor of ${\sim}\:{\times 1.33}$.

\begin{figure}[t!]
\centering
\includegraphics[width=0.48\textwidth]{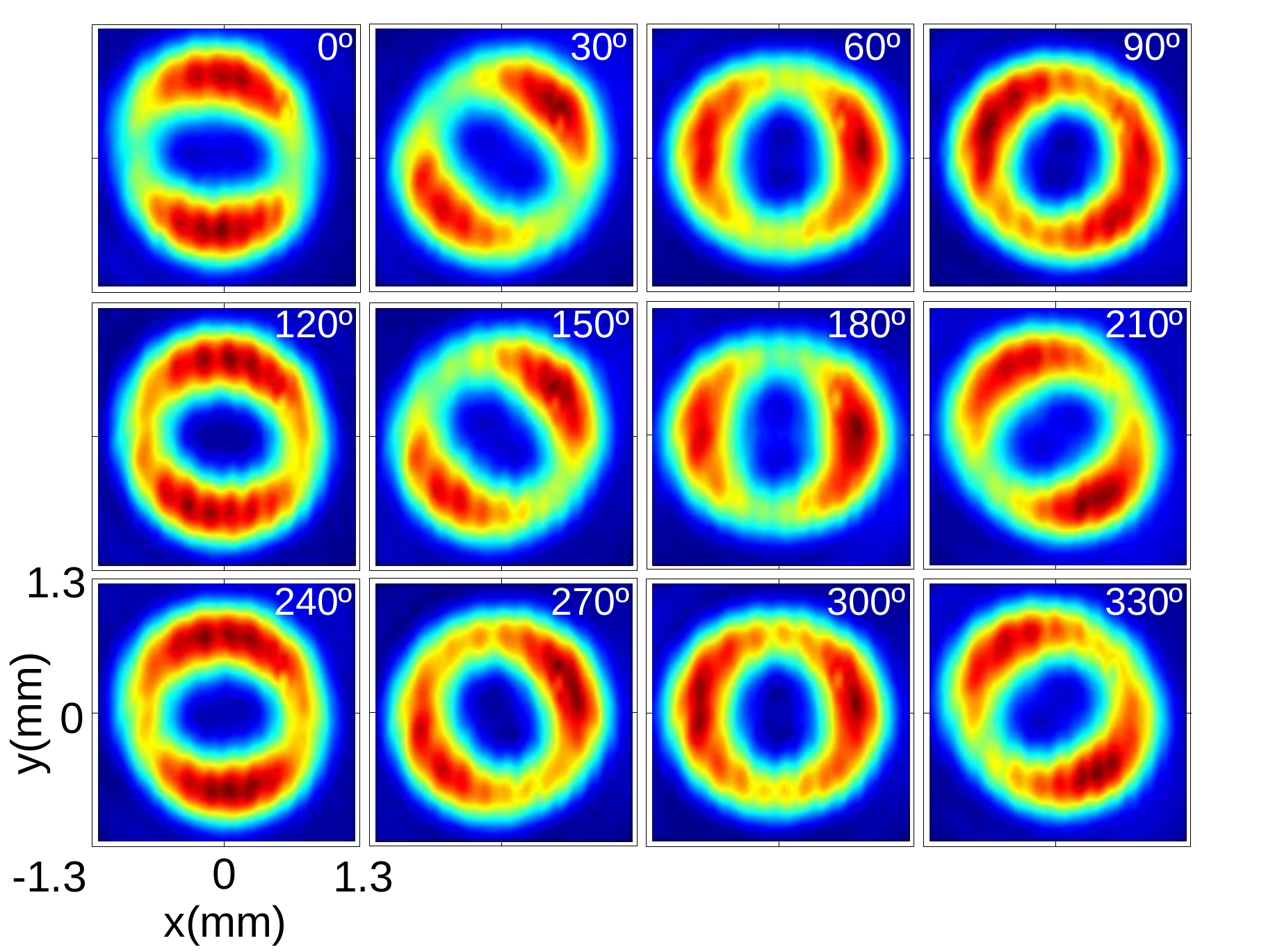}
\figurecaption{Observed interference patterns}{
Representative interference patterns observed with $m_2=-2$ for various values of the polarizer angle $\theta$.
}
\label{fig-results-rotating-interference}
\end{figure}

\begin{figure}[t!]
\vspace{0.5em}
\centering
\setlength{\tabcolsep}{0mm}
\renewcommand{\arraystretch}{0.}
\newcommand{\vortexFigScale}{0.74}
\begin{tabular}{rl}
\includegraphics[scale=\vortexFigScale]{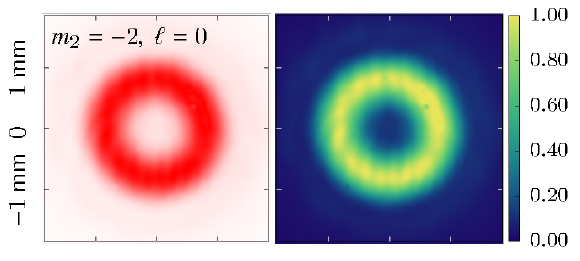} &
\includegraphics[scale=\vortexFigScale]{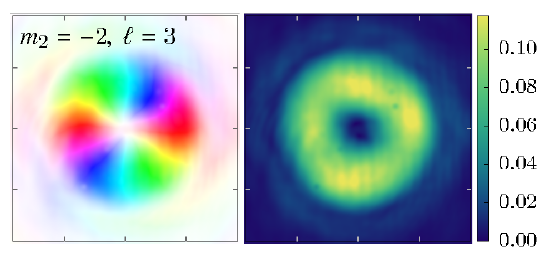} \\[-1mm]
\includegraphics[scale=\vortexFigScale]{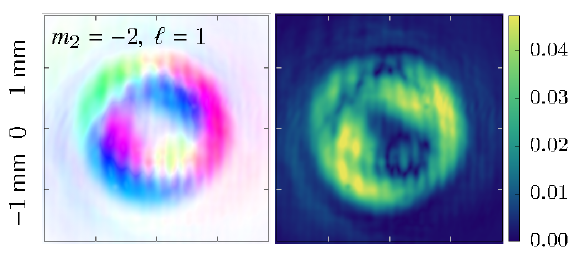} &
\includegraphics[scale=\vortexFigScale]{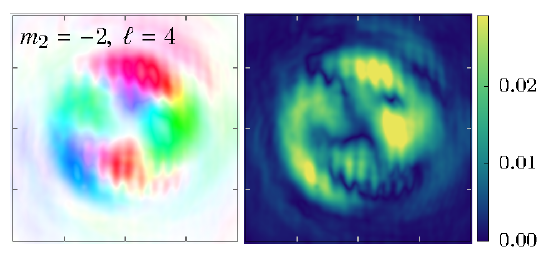} \\[-1mm]
\includegraphics[scale=\vortexFigScale]{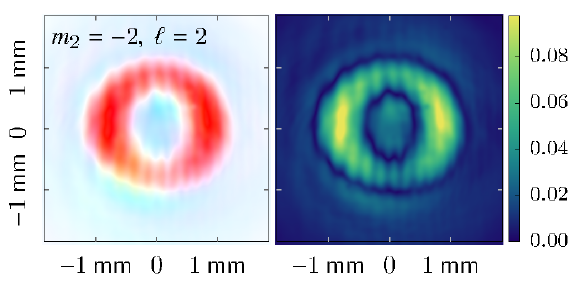} &
\includegraphics[scale=\vortexFigScale]{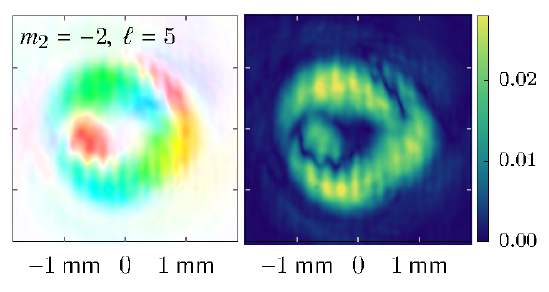}
\end{tabular}
\figurecaption{Transformed experimental results}{
Fourier transform of the experimental results of \reffig{fig-results-rotating-interference}, shown in both magnitude (right-hand panels) and phase (left-hand panels; $\arg(\tilde f_\ell(x,y))$ as the hue and $|\tilde f_\ell(x,y)|$ as the saturation).
We normalize $|\tilde f_\ell(x,y)|$ to its maximum value within each plot, but we show its scale with respect to the base $\ell=0$ intensity $|\tilde f_0(x,y)|$ on the density-plot tick marks. 
Thus, the only meaningful channel $\ell>0$ is $\ell=3$, with a dominant signal and a clear phase winding.
}
\label{fig-vortex-analysis2}
\end{figure}

In order to turn this beam into a high-quality higher-order Laguerre-Gauss Beam (LGB) we used a Spatial Light Modulator (SLM, Hamamatsu X10468-2, $792\times 600$ pixels with pixel pitch of $\SI{20}{\micro m}$) so as to modulate phase of the incoming beam by phase patterns for LGB displayed on the SLM. 
A half-wave plate HWP was used to change the polarization orientation of the beam to horizontal, as required by the SLM, and another HWP was used return the reflected beam to vertical polarization. The output beam is a high-quality LGB with vertical polarization, with a topological charge $m_2$ controlled by the phase winding on the~SLM.

Following the dichroic mirror DM, the beam centered at $\SI{1550}{nm}$ was reflected and propagated along the arm A2. First it was transmitted through a lens $L_{3}$ with focal length of $f=\SI{15}{cm}$ which forms a telescope with the lens $L_{1}$ with magnification $\times 1$ to collimate the beam.

In order to obtain counter-rotating circular polarizations on both beams, a quarter-wave plate was placed in each arm, so that along the arm A1 and just after the SLM our LGB was transmitted through the quarter-wave plate QWP1 oriented at $\SI{45}{\degree}$, also along the arm A2 and just after the DM the beam with wavelength centered at $\SI{1550}{nm}$ was transmitted through the quarter-wave plate QWP2 oriented at $\SI{-45}{\degree}$.

\begin{figure}[t!]
\centering
\setlength{\tabcolsep}{0.mm}
\renewcommand{\arraystretch}{0.}
\newcommand{\vortexFigScale}{0.74}
\begin{tabular}{rl}
\includegraphics[scale=\vortexFigScale]{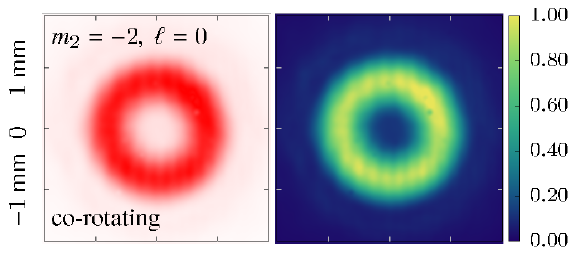} &
\includegraphics[scale=\vortexFigScale]{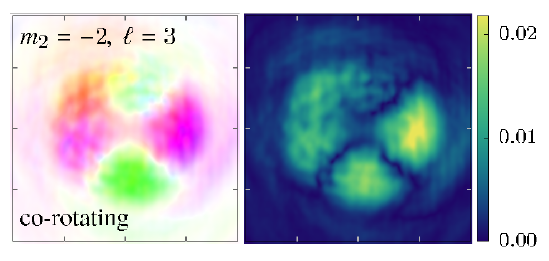} \\[0mm]
\includegraphics[scale=\vortexFigScale]{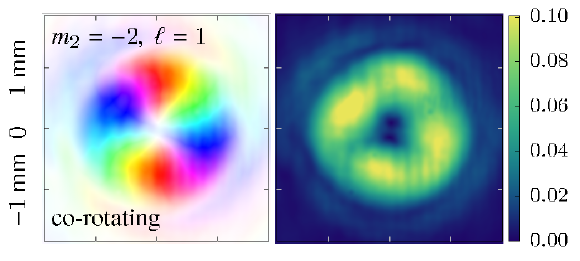} &
\includegraphics[scale=\vortexFigScale]{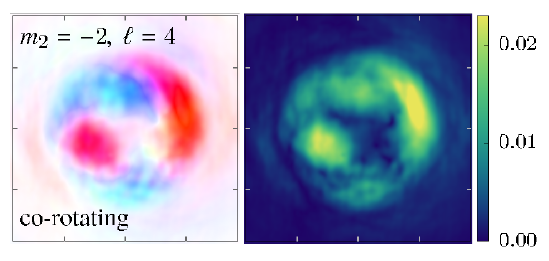} \\[0mm]
\includegraphics[scale=\vortexFigScale]{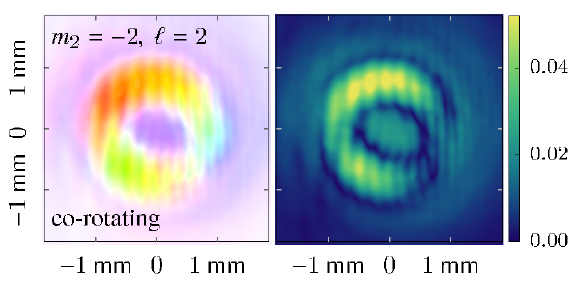} &
\includegraphics[scale=\vortexFigScale]{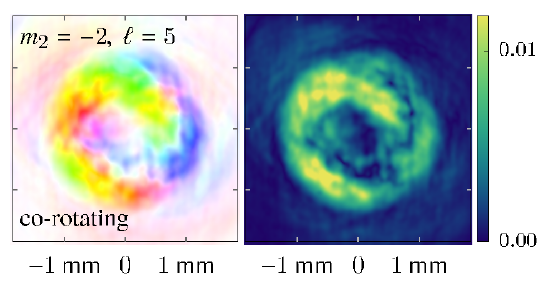}
\end{tabular}
\figurecaption{Fourier analysis for co-rotating polarizations}{
Analysis of the results, as in \reffig{fig-vortex-analysis2}, for co-rotating circular polarizations on both arms of the interferometer, with $m_2=-2$. Here the dominant component shifts from $\ell=3$ to $\ell=1$, showing a nonzero phase winding on $T_{1,3}$, and with that a `true' vector vortex.
}
\label{fig-vortex-analysis-co-rotating}
\end{figure}

\begin{figure}[t!]
\vspace{0.5em}
\centering
\setlength{\tabcolsep}{0.2mm}
\newcommand{\phVorScale}{0.64}
\begin{tabular}{cccc}
\includegraphics[scale=\phVorScale]{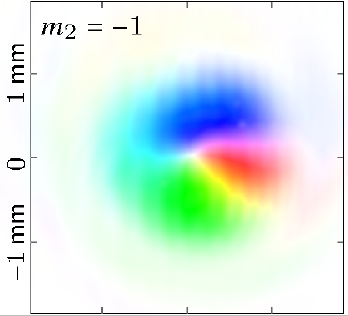} &
\includegraphics[scale=\phVorScale]{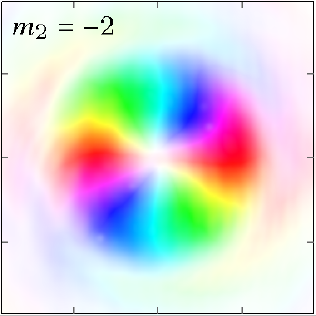} &
\includegraphics[scale=\phVorScale]{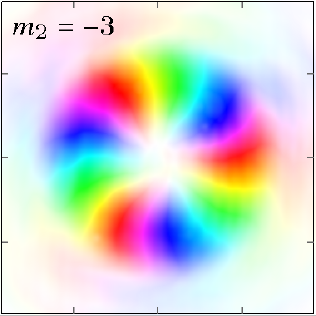} &
\includegraphics[scale=\phVorScale]{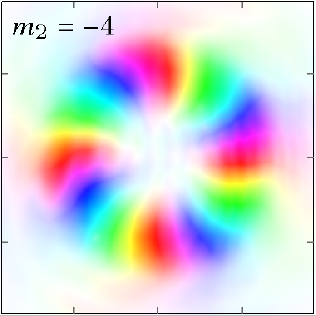} \\[-0.5mm]
\includegraphics[scale=\phVorScale]{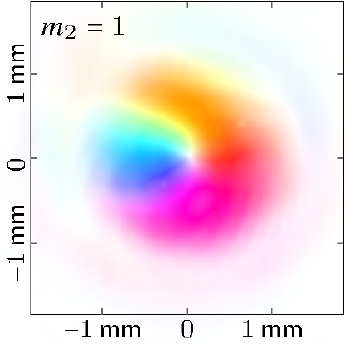} &
\includegraphics[scale=\phVorScale]{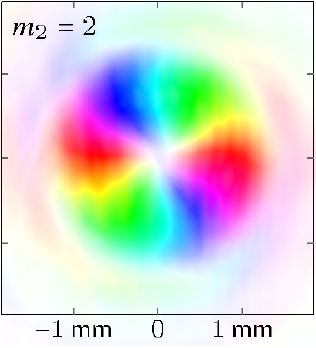} &
\includegraphics[scale=\phVorScale]{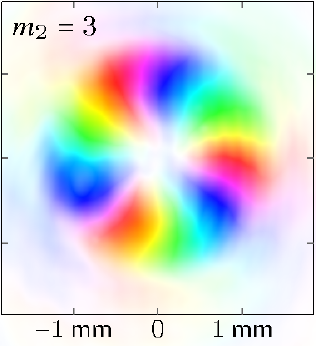} &
\includegraphics[scale=\phVorScale]{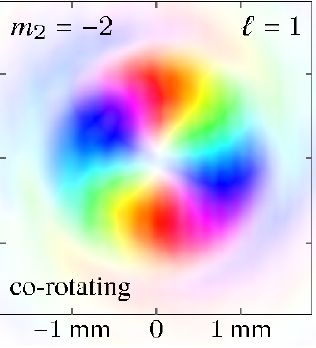}
\end{tabular}
\figurecaption{Observed phase vortices}{
Experimentally-ob\-ser\-ved phase vortices, in $T_{3,3}$ plotted as in Figs.~\ref{fig-setup-and-results} and~\ref{fig-vortex-analysis2}, for different values of the second harmonic's OAM $m_2$, as well as for the co-rotating case of \reffig{fig-vortex-analysis-co-rotating}.
}
\label{phasevortices}
\end{figure}

\begin{figure}[b!]
\centering
\begin{tabular}{l}
\includegraphics[height=0.14\textwidth]{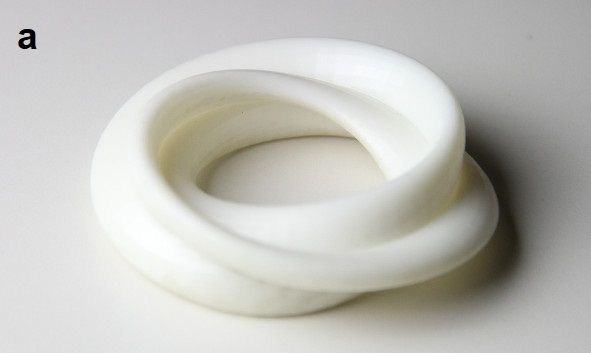}
\hspace{0.2mm}
\includegraphics[height=0.14\textwidth]{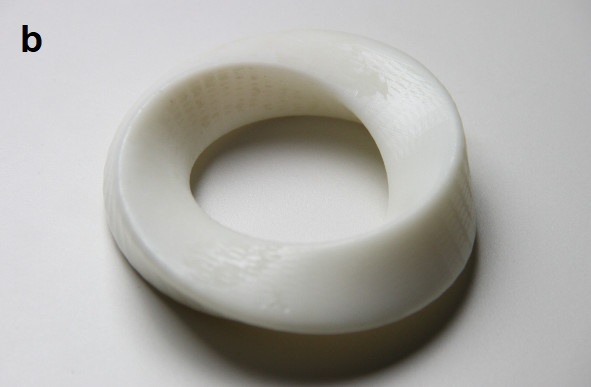}\\
\includegraphics[height=0.14715\textwidth]{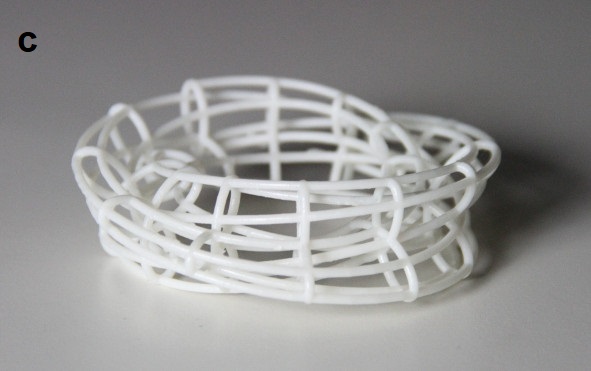} 
\end{tabular}
\figurecaption{3D prints of the torus-knot topology}{%
Models of the torus-knot beam topology of \reffig{fig-knot-buildup}, available at \citer{supplementary-information}, printed in 3D using resin-based stereolithography on a Formlabs Form 2 printer. 
3D Prints: Xavier Menino (ICFO); 
Image Credits: {\textcopyright}\,ICFO.
}
\label{fig-3d-print-photos}
\end{figure}

For the detection stage, both beams are transmitted through a linear polarizer LP1 and LP2, for each arm A1 and A2 respectively, to project the polarization state at angle $\theta$. We keep the detection nonlinear crystal (NLC2, 10-mm-long PPLN) stationary, to ensure stable phase-matching, rotating the polarization after LP1 and LP2 into its phase-matching angle using half-wave plates HWP1 and HWP2 at angles $\theta/2$. The NLC2 crystal was placed in the arm A2 between the lenses $L_{4}$ and $L_{5}$, with focal lengths of $f = \SI{15}{cm}$ and $f = \SI{30}{cm}$ respectively, so as to magnify the beam by a factor of ${\sim}\:{\times 2}$. Pump light was eliminated using a bandpass filter (F) centered at $\SI{775}{nm}$ with a bandwidth of $\SI{10}{nm}$. The two beams were recombined by a beamsplitter and their interference pattern was observed with a CCD camera (resolution of $1200 \times 1600$  pixels, pixel width $\SI{4.4}{\micro m}$).

We recorded data by imaging the interference patterns produced by setting the linear polarizer angles $\theta$ (and with them the half-wave plate angles $\theta/2$) between $\theta = \SI{0}{\degree}$ and $\theta = \SI{330}{\degree}$ in steps of $\SI{30}{\degree}$, which produces a rigid rotation on the interference pattern as shown in \reffig{fig-results-rotating-interference}. To complete the polarization tomography, we Fourier transform the $N=12$ measured images $f_{\theta_j}(x,y)$, over $\theta_j=2\pi j/N$ with $j=0,1,\ldots,N-1$, to $\tilde f_\ell(x,y) = \sum_{j=0}^{N-1}e^{i\ell \theta_j}f_{\theta_j}(x,y)$; we show a representative example in \reffig{fig-vortex-analysis2}. Generally, there is some residual signal on most components, but only the dominant channel at $\ell=3$ carries a nonzero phase winding.

A similar analysis carried out with co-rotating polarizations on both arms (achieved by setting QWP2 to $+\SI{45}{\degree}$), shown in \reffig{fig-vortex-analysis-co-rotating}, exhibits a shift in the domminant Fourier component from $\ell=3$ to $\ell=1$, with a nonzero phase winding on the latter, corresponding to the observation of a phase singularity in the dipole field moment $T_{1,3}$.

\section{Data and code availability}
The code and data used for this paper have been archived in Zenodo at \href{https://doi.org/10.5281/zenodo.2649391}{doi:\allowbreak{}10.5281/\allowbreak{}zenodo.\allowbreak{}2649391}.

\section{Supplementary material}
We include as supplementary material, available as  \citer{supplementary-information} at \href{https://doi.org/10.5281/zenodo.2597667}{doi:\allowbreak{}10.5281/\allowbreak{}zenodo.\allowbreak{}2597667}, 3D-printable models of figures~\ref{fig-twisted-trefoil-knot} and \ref{fig-initial-knot}. We include here, in \reffig{fig-3d-print-photos}, photographs of the finished models.

\end{document}